\documentclass[10pt]{article}
\usepackage{fancyhdr}
\usepackage{extramarks}
\usepackage{amsmath}
\usepackage{amsthm}
\usepackage{amsfonts}
\usepackage{siunitx}
\usepackage{tikz}
\usepackage[plain]{algorithm}
\usepackage{algpseudocode}
\usepackage{multirow}
\usepackage{booktabs}
\usepackage{graphicx}
\usepackage{subfigure}
\usepackage[colorlinks,linkcolor=black,anchorcolor=black,citecolor=black,urlcolor=blue]{hyperref}
\usepackage{amsmath,bm}
\usepackage{booktabs}
\usepackage{mathtools}
\usepackage{amssymb}
\usepackage{caption}
\usepackage{capt-of}
\usepackage{mciteplus}
\usepackage{cite}
\usepackage{mathrsfs}
\usepackage[title,titletoc,toc]{appendix}

\usepackage{xr}
\usepackage{xr-hyper}
\usepackage{hyperref}

\usepackage{parskip}
\usepackage{soul}
\usepackage{textcomp}
\usepackage[colaction]{multicol}
\usepackage[switch]{lineno}
\usepackage{lipsum}
\usepackage{etoolbox}
\usepackage{longtable}
\usepackage{array}
\usepackage{tablefootnote}
\usepackage{ragged2e}
\newcolumntype{C}[1]{>{\centering\arraybackslash}p{#1}}
\usetikzlibrary{automata,positioning}
\usepackage{adjustbox}
\usepackage{makecell}
\usepackage{pdflscape}
\usepackage{setspace}
\usepackage{color}
\usepackage{authblk}

\DeclareMathOperator{\im}{im}

\usetikzlibrary{automata,positioning}

\makeatletter
    \newcommand*{\addFileDependency}[1]{
    \typeout{(#1)}
    \@addtofilelist{#1}
    \IfFileExists{#1}{}{\typeout{No file #1.}}
    }

\makeatother

\usepackage{url}

\begin{document}

\title{Machine learning predictions from unpredictable chaos}
\author[1,2,*]{Jian Jiang}
\author[1]{Long Chen}
\author[1]{Lu ke}
\author[1]{Bozheng Dou}
\author[1]{Yueying Zhu}
\author[1]{Yazhou Shi}
\author[1]{Huahai Qiu}
\author[1]{Bengong Zhang}
\author[3]{Tianshou Zhou}
\author[2,4,5,+]{Guo-Wei Wei}

\affil[1]{Research Center of Nonlinear Science, School of Mathematical and Physical Sciences, Wuhan Textile University, Wuhan, 430200, P R. China}
\affil[2]{Department of Mathematics, Michigan State University, East Lansing, Michigan 48824, USA}
\affil[3]{Key Laboratory of Computational Mathematics, Guangdong Province, and School of Mathematics, Sun Yat-sen University, Guangzhou, 510006, P R. China}
\affil[4]{Department of Electrical and Computer Engineering Michigan State University, East Lansing, Michigan 48824, USA}
\affil[5]{Department of Biochemistry and Molecular Biology Michigan State University, East Lansing, Michigan 48824, USA}
\affil[*]{Corresponding author: jjiang@wtu.edu.cn}
\affil[+]{Corresponding author: weig@msu.edu}

\date{} 

\maketitle

\begin{abstract}
Chaos is omnipresent in nature, and its understanding provides enormous social and economic benefits. However, the unpredictability of chaotic systems is a textbook concept due to their sensitivity to initial conditions, aperiodic behavior, fractal dimensions, nonlinearity, and strange attractors. In this work, we introduce, for the first time, chaotic learning, a novel multiscale topological paradigm that enables accurate predictions from chaotic systems. We show that seemingly random and unpredictable chaotic dynamics counterintuitively offer unprecedented quantitative predictions. Specifically, we devise multiscale topological Laplacians to embed real-world data into a family of interactive chaotic dynamical systems, modulate their dynamical behaviors, and enable the accurate prediction of the input data. As a proof of concept, we consider 28 datasets from four categories of realistic problems: 10 brain waves, four benchmark protein datasets, 13 single-cell RNA sequencing datasets, and an image dataset, as well as two distinct chaotic dynamical systems, namely the Lorenz and Rossler attractors. We demonstrate chaotic learning predictions of the physical properties from chaos. Our new chaotic learning paradigm profoundly changes the textbook perception of chaos and bridges topology, chaos, and learning for the first time.
\end{abstract}
 

\maketitle

\newpage

\section{Introduction} 

Chaos in dynamical systems refers to the complex behavior that emerges from deterministic, yet sensitive complex systems such as double pendulum, social network, turbulent fluid flow,  weather systems, etc \cite{aguirre2009modeling}. Characterized by a sensitive dependence on initial conditions and aperiodic behavior, chaos challenges the ability to predict long-term behavior \cite{crutchfield2012between}. Nonlinearity is another key feature of chaos, where interactions between system elements lead to unpredictable and often erratic dynamics. The seminal work of Edward Lorenz in the 1960s \cite{lorenz1963deterministic}, particularly his study of the weather model, introduced the notion that chaotic systems are fundamentally unpredictable over long timescales. This insight laid the foundation for chaos theory, marking a shift in the way scientists viewed complex systems. Over half a century, the unpredictability of chaotic systems has been a textbook concept in both theoretical and applied dynamics, shaping research and application in various fields, from meteorology to engineering \cite{boccaletti2000control}. 


Chaos theory and chaotic dynamics have profoundly impacted numerous disciplines by providing a framework for understanding complex and nonlinear phenomena\cite{li1975period}. For example, in meteorology, chaos theory revolutionized weather prediction, highlighting the challenge of long-term forecasts due to sensitive dependence on initial conditions \cite{selvam2017nonlinear}. In engineering, chaotic systems have inspired the design of robust control strategies and fault-tolerant systems, especially in nonlinear circuits and mechanical systems \cite{dong2022neuromorphic}. In biology, the study of chaos has shed light on irregular patterns in physiological processes, such as cardiac rhythms and neural activity, enhancing the understanding of health and disease dynamics \cite{heltberg2021tale}. Similarly, in economics, chaotic models have been employed to explain irregular market fluctuations, providing insights into complex financial systems \cite{fernandez2023overview}. The unpredictability inherent in chaos also plays a pivotal role in encryption \cite{zhang2023chaos} and secure communications \cite{zaher2011design}, where it ensures high levels of randomness and security. Across these fields, chaos theory has underscored the importance of recognizing and leveraging unpredictability to uncover hidden structures and drive innovation from complex systems.

Recently,  machine learning (ML) has exploited the deterministic nature of chaos theory to predict certain properties of chaotic systems, such as spatial extent and attractor dimension \cite{pathak2018model}, fluctuations in a chaotic cancer model \cite{ramadevi2022chaotic}, and the evolution of chaotic dynamical systems \cite{fan2020long}. Deep learning (DL) methods have also been utilized to demonstrate chaotic synchronization \cite{weng2022modeling}. 
 However, utilizing chaotic systems for quantitative prediction remains an unthinkable challenge because of their unpredictability.    

In this work, we propose chaotic learning to transform unpredictable chaotic systems into accurate predictors. Our approach utilizes a novel multiscale topological method, known as persistent Laplacian (PL) \cite{wang2020persistent}, to embed realistic data in a family of heterogeneous chaotic systems. More specifically, each data point is modeled as a chaotic oscillator, and its interactions with other oscillators are captured with multiscale Laplacians.  The resulting data-embedded chaotic systems exhibit rich chaotic patterns that faithfully manifest the traits of the original data, leading to  accurate ML predictions. The proposed topology-enabled predictions from chaos (TEPC) utilizes various chaotic dynamics, such as the Lorenz and Rossler systems, for chaotic learning. 
Additionally, TEPC may employ different persistent topological Laplacians, including persistent spectral graph \cite{wang2020persistent} or persistent sheaf Laplacian \cite{wei2025persistent}. 
Finally, TEPC has been extensively validated by using  a wide variety of benchmark data, including organ-scale brain electroencephalogram (EEG) data classification, molecular-scale protein B-factor prediction, cellular-scale single-cell RNA sequence data classification, and general image classification problems. TEPC turns chaotic theory on its head, challenging long-held perception of the unpredictability of chaos. It also bridges the gap between traditional chaos and modern learning algorithms.

\section{Results}
\subsection{Overview of chaotic learning --- topology-enabled predictions from chaos (TEPC)}

\begin{figure}[!tpb]
\centering
\includegraphics[width=13.5cm]{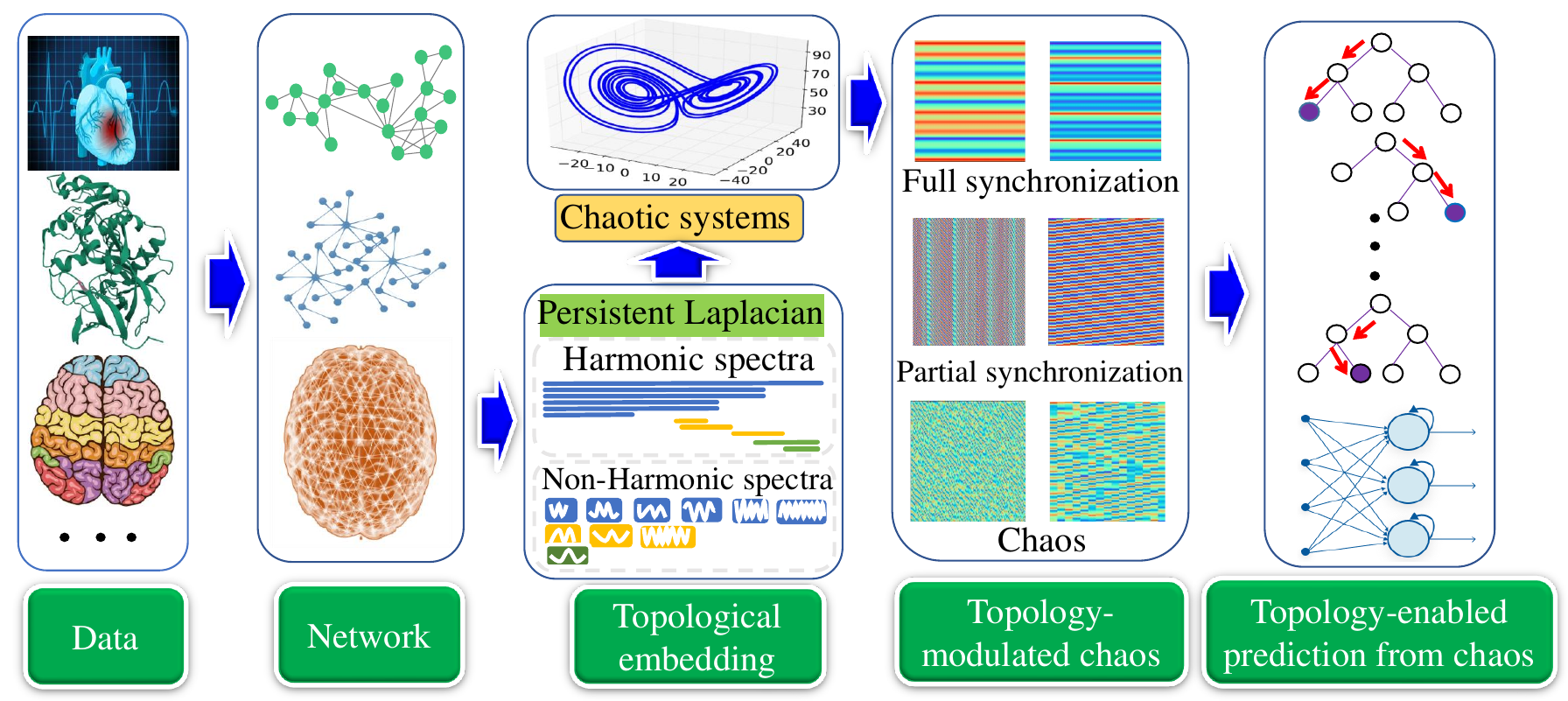}
\caption{Illustration of the TEPC platform, which provides a general framework to predict the  properties of the data of interest, by means of algebraic topology, chaos theory, and  machine learning (ML), including deep learning (DL). First, real-world data is represented by networks where nodes denote individual data points (i.e., atoms, cells, brain nodes, images, etc.) and the links indicate the interaction or the connectivity between nodes. Each node is injected with a chaotic dynamics, like the Lorenz and Rollser oscillator. Then, topological Laplacians, such as persistent Laplacian or persistent sheaf Laplacian, are applied to the networks to encode the multiscale connectivity matrices among nodes, leading to  a family of chaotic systems. Subsequently, the evolution  trajectories of the topology-regulated chaotic systems are extracted and their statistical features are obtained for various downstream ML/DL tasks. Alternatively, chaotic trajectories are seamlessly fed into a natural language processing (NLP) model for predictions.  
The brain and heart pictures were adopted from \url{https://www.freepik.com.}}
\label{framework}
\end{figure}

Fig. \ref{framework} illustrates the TEPC platform that provides a general workflow of topology-enabled ML predictions from chaotic dynamics. First, the data of interest are collected from real world problems, and then an interaction network or a hypergraph is constructed where nodes are individual data points, and the links represent the correlations or connections between nodes. Next, each node is introduced with a chaotic dynamics, such as the Lorenz and Rossler dynamics \cite{osipov1997phase}. A family of chaotic systems is constructed by using topological Laplacians  \cite{wang2020persistent}, which encode the interactions into the multiscale connectivity matrices for chaotic systems. 
Subsequently, the multiscale evolution trajectories of the chaotic systems, which may display various patterns, such as full synchronized, partial synchronized, and unsynchronized chaos, faithfully embed the property of the data of interest. Finally, statistic features are extracted from chaotic trajectories and utilized for downstream ML or DL predictions. Alternatively, chaotic trajectories  choice is that these chaotic dynamics is fed into recurrent neural network (RNN) or long short-term memory (LSTM) for downstream predictions as discussed in Section S2 of the Supporting Information. 

The proposed chaotic learning is an ML algorithm for real-world data prediction rather than a new model for chaos or nonlinear dynamics. Data-embedded chaotic systems have heterogeneous nodal dynamics. Even if complete synchronizations may eventually achieved, the heterogeneous synchronization process of individual nodes reveals the intrinsic properties of individual nodes, which forms the foundation of the proposed chaotic learning model.

\subsection{Illustration of chaotic learning principle} 

\begin{figure}[!tpb]
\centering
\includegraphics[width=16.6cm]{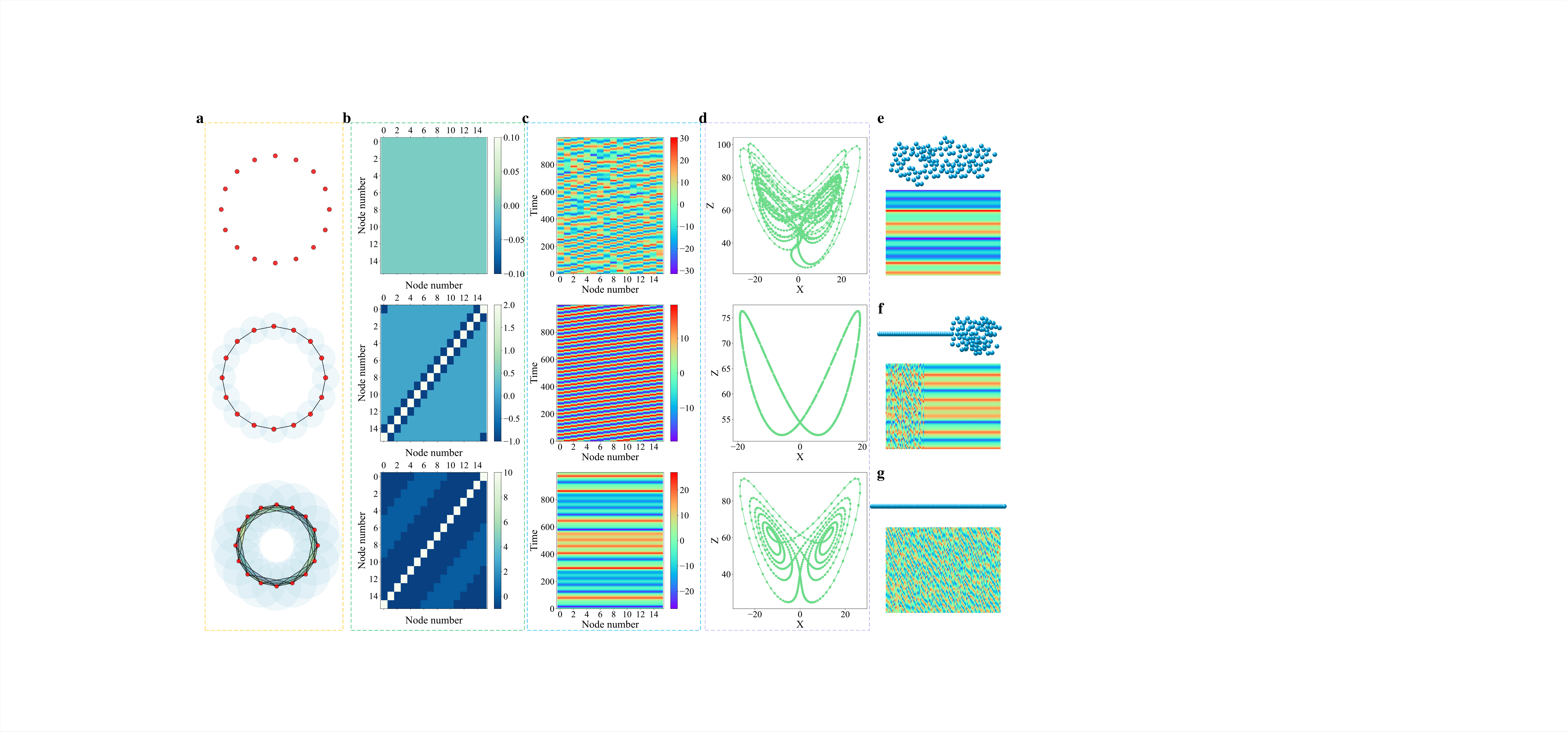}
\caption{Illustration of persistent Laplacian induced transition from chaos to periodicity. 
\textbf{a}: Each of the sixteen nodes in a regular hexadecagon is injected with a Lorenz oscillator. The  coupling or connectivity  of these nodes are given by the filtration radius of persistent Laplacian. Three typical filtration patterns are displayed in three charts.   
\textbf{b}: The connectivity matrices for the coupled chaotic systems are generated by persistent Laplacian at three distinct filtration radii shown in three charts.  
\textbf{c}: The trajectories of the coupled chaotic oscillators with random initial values, showing chaos (top), partial synchronization (middle), and full synchronization (bottom). 
\textbf{d}: Irregular orbit, periodic orbit, and butterfly wing patterns for three chaotic systems.
\textbf{e}: The folded geometry and its synchronous dynamics of a 120-element point cloud.    
\textbf{f}: The partially folded geometry and its partially synchronized  dynamics of a 120-element point cloud. 
\textbf{g}: The unfolded geometry and its chaotic dynamics of a 120-element point cloud. 
}
\label{polygon16}
\end{figure}

To illustrate the working principle of the proposed chaotic learning paradigm,  let us consider two simple models. Our first model is a simple model with sixteen nodes as shown in Fig. \ref{polygon16}{\bf a}  to demonstrate the proposed paradigm. The three-dimension coordinates of each node are given in Table S13 of the Supporting Information. Each point or node can be regarded as a data point or an individual object, such as an atom in a molecule,  a protein in a protein-protein network, or a neuron in a brain network. To extract the localized properties of the sixteen nodes, we introduce a chaotic oscillator, or a nonlinear chaotic dynamic to each node. More specifically, we utilize the Lorenz oscillator although any other chaotic oscillators can be used as well. 
The persistent Laplacians are constructed for the set of oscillators, resulting in a family of connectivity matrices for the coupled chaotic systems.  Three typical connectivity matrices are given in Fig. \ref{polygon16}{\bf b},  corresponding to three different filtration radius given in Fig. \ref{polygon16}{\bf  a}. 
The details of these chaotic systems  can be found in Section S1 of the Supporting Information. Initial values for all oscillators are randomly chosen.  The smaller filtration radius, the narrower coupling among oscillators is. For example, with the filtration radius shown at the top of Fig. \ref{polygon16}{\bf a}, as there is no overlap between the nodes, Laplacian is zero everywhere as shown at the top of Fig. \ref{polygon16}{\bf b}. Consequently, we have a set of isolated chaotic dynamics as shown on the top of Fig. \ref{polygon16}{\bf c}, resulting in irregular orbits on the top of Fig. \ref{polygon16}{\bf d}.

As the filtration radius increases showing in the middle of Fig. \ref{polygon16}{\bf a}, there are some interactions in the system given by the Laplacian in the middle of Fig. \ref{polygon16}{\bf b}, which causes partial synchronization among chaotic oscillators during the temporal evolution of system (see the middle of Fig. \ref{polygon16}{\bf c}) and periodic orbit in the middle of Fig. \ref{polygon16}{\bf d}. When the filtration radius continuously increases, the coupling among the nodes becomes sufficiently strong to create a full synchronization in the system, and each oscillator resembles the well-known wings of butterfly as plotted at the bottom of Fig. \ref{polygon16}{\bf d}. In this simple model, it is found that multiscale topological Laplacians induce the transition from chaos to periodicity. However, since the geometry of the above example is very regular and thus, their dynamics are  
homogeneous as shown in Figs. \ref{polygon16}{\bf a-d}.  

In our second model, we consider three sets of irregular geometries as shown in Figs. \ref{polygon16}{\bf e-g} to illustrate how irregular geometries induce heterogeneous chaotic dynamics, including partially synchronized chaos.  We create a set of 120 nodes as shown on the top of each chart. The detailed coordinates are given in Tables S14, S15, S16 of the Supporting Information, respectively. The geometric shapes of these nodes are schematically displayed in Figs. \ref{polygon16}{\bf e-g}. We inject the Lorenz dynamics into each of these nodes with the same parameters as those in the early example. However, for chaotic systems, we set coupling strength as $\epsilon=$ 1.1, 1.0, and 0.1 for Figs. \ref{polygon16}{\bf e-g}, respectively. The connectivity matrices are generated by using the persistent Laplacian algorithm with  the filtration radii being 1, 1, and 0.2 for the point cloud in Figs. \ref{polygon16}{\bf e-g}, respectively. We used the forward Euler scheme with the time increment of $h=10^{-3}$  for the time integration. All details of parameters used in Fig. \ref{polygon16} can be found in Table S1 of the Supporting Information. 

The first set of 120 nodes is folded, which leads to a fully synchronized chaotic system as shown in Fig. \ref{polygon16}{\bf e}. This synchronous chaos is very similar to that in the bottom of Fig. \ref{polygon16}{\bf c}. In Fig.  \ref{polygon16}{\bf f}, the partially folded conformation having 40 unfolded nodes and 80 folded ones, which gives rise to a partially synchronized dynamic behavior.   In Fig. \ref{polygon16}{\bf g}, the dynamics of the completely unfolded conformation is completely chaotic. 
It is this strong geometry-dynamics correlation as well as the rich behavior in chaos synchronization that lays the foundation to embed a wide variety of real-world data into chaotic systems and extract the full set of seemingly chaotic information for the accurate prediction of the underlying data.   

Although the filtration radius can be regarded as a threshold used to define a family of networks, it is not highly sensitive across different applications. This is because we implement a topological filtration procedure on the weighted Laplacian matrix to construct a family of subgraphs or networks, which encode the interactions between oscillators into the multiscale connectivity of the networks. The resulting multiscale evolution trajectories of these networks exhibit various patterns, such as synchronized, partially synchronized, and unsynchronized chaos, that effectively capture intrinsic traits of the data for downstream ML predictions. Consequently, in our model, the threshold does not require fine-tuning; an "educated choice" is sufficient to achieve good predictive performance. 
 
\subsection{Chaotic learning prediction of brain EEG data } \label{sec:maintext_EEG}

 To present a proof-of-principle exploration of the proposed chaotic learning paradigm and demonstrate its predictive power, we consider brain electroencephalogram (EEG) datasets.  The EEG is obtained from a medical measurement of brain electrical activity using a number of electrodes applied to the scalp. It plays a significant role in the diagnosis of various patient conditions, such as epilepsy, schizophreniacs, sleep disorders, and brain tumors. In our analysis, the EEG data have 300 signals and were collected from 5 epilepsy patients and 5 healthy individuals, classified into three distinct categories: normal, preictal, and seizure \cite{andrzejak2001indications}. Each category contains 100 single-channel EEG signals. This data set can be accessed from the official website of the Department of Epileptology of the University of Bonn (\url{https://www.ukbonn.de/epileptologie/arbeitsgruppen/ag-lehnertz-neurophysik/downloads/}). Their details can be found in Section S5 of the Supporting Information. 

 We embed the real brain activity time series data into nonlinear dynamics systems by considering each EEG signal as a nonlinear oscillator. We further compute the mean coupling matrix of this brain neural network by calculating the Pearson correlation coefficient between each pair of 300 signals, respectively. We then apply persistent Laplacian to obtain a family of connectivity matrices of brain neural networks with the filtration process. For each connectivity matrix, we introduce a chaotic system and collect its trajectories. Their statistical values are computed for EEG classification. We use ten-fold cross-validation with 10 random seeds and k-nearest neighbor classifier (KNN) algorithm in the prediction. The details of parameters used in the implementation are given in  the Supporting Information. To demonstrate the predictive power of TEPC, we consider  classification predictions of the EEG data. Our TEPC was implemented with two different oscillators i.e., Lorenz and Rossler and two neural network models i.e., RNN and LSTM, denoted as TEPC (Lorenz), TEPC (Rossler), TEPC (RNN), and TEPC (LSTM), respectively. 
 Their comparison with other existing methods is shown in Fig. \ref{results}a, including Fuzzy sugeno1 \cite{acharya2012application}, Random forest \cite{bhattacharyya2017multivariate}, PCA \cite{ghosh2007mixed}, SVM5 \cite{bhattacharyya2017tunable}, Fuzzy sugeno2 \cite{acharya2012automated}, LMB neural network \cite{ghosh2009new}, Fuzzy sugeno3 \cite{acharya2012use}, SVM2 \cite{acharya2011automatic}, SVM4 \cite{ur2013automated}, SVM3 \cite{acharya2011application}, C4.5 decision tree \cite{martis2012application}, Multi-spiking \cite{ghosh2009spiking}, KNN \cite{guo2011automatic}, SVM1 \cite{faust2010automatic}, GMM2 \cite{chua2009automatic}, GMM3 \cite{chua2011application}, Spiking \cite{ghosh2007improved}, and CNN \cite{acharya2018deep}. The ten-fold cross validation is employed in all predictions. ACC, SEN, and SPEC in Fig. \ref{results}a are accuracy, sensitivity, and specificity, respectively. 
From Fig. \ref{results}a, we can find that TEPC (Lorenz), TEPC (Rossler), TEPC (RNN), and TEPC (LSTM) obtains all 100\% on these three metrics, which are equivalent or higher than those of other existing methods. The confusion matrices for all ten folds of these results are given in Tables S8, S9, S10, and S11 of the Supporting Information. These results indicate that TEPC is a general approach that is robust with respect to the selection of dynamic models and ML algorithms.

\subsection{Predicting protein B-factors from chaotic systems}

\begin{figure}[!tpb]
\centering
\includegraphics[width=16.5cm]{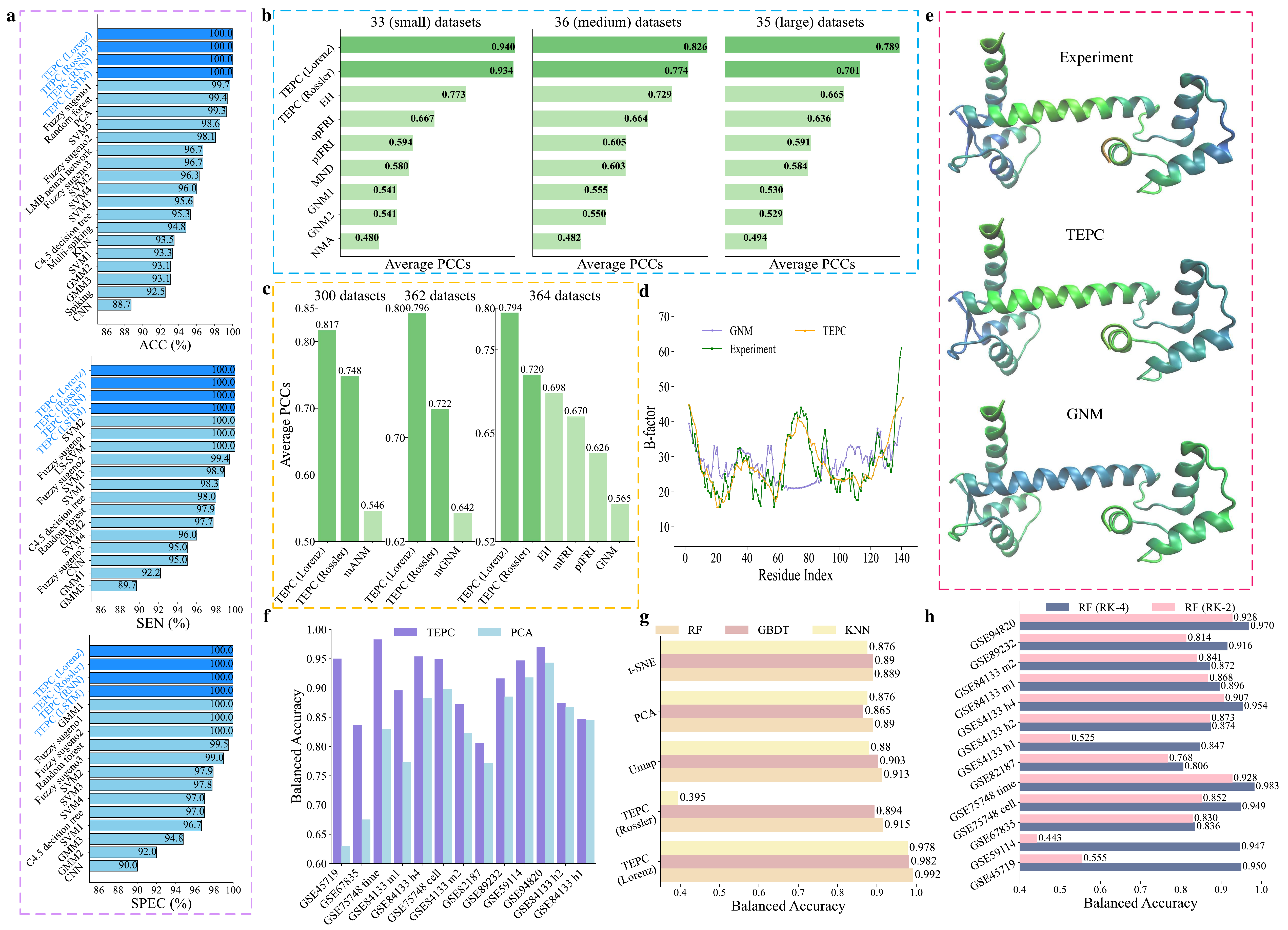}
\caption{The comparison of prediction results between TEPC method and other existing methods on different types of data, including brain EEG signals, B-factors of proteins, single cell RNA sequences, and image classification. \textbf{a}: The comparison of brain EEG data classification among TEPC methods (i.e., TEPC (Lorenz), TEPC (Rossler), TEPC (RNN), and TEPC (LSTM)) and other existing methods. The classification involves three categories:  normal, preictal, and seizure. ACC, SEN, and SPEC refer to the accuracy, sensitivity, and specificity, respectively. \textbf{b}-\textbf{c}: The comparison of B-factor predictions between our TEPC methods with Lorenz and Rossler oscillators and other literature methods on three datasets, on a benchmark dataset of 300, 362, and 364 proteins. \textbf{d}: The comparison of TEPC, GNM prediction value and experimental value for each residue. The $x$-axis means the residue index, and the $y$-axis means the B-factor.  \textbf{e}: The three-dimension (3D) structures of protein calmodulin (PDB ID: 1CLL), colored by experimental B-factors, TEPC predicted B-factors, and GNM method predicted B-factors from up to bottom, respectively.  \textbf{f}: The comparison of the balanced accuracy between our TEPC method with Lorenz oscillators and principal components analysis (PCA) method through a random forest (RF) classifier on 13 single cell RNA sequence data. \textbf{g}: The comparison of balanced accuracy between our TEPC method with Lorenz and Rossler oscillators and UMAP, PCA, t-SNE method with RF, gradient boosting decision tree (GBDT), and k-nearest neighbor classifier (KNN) classifier on Coil-20 image data. \textbf{h}: The comparison of balanced accuracy between fourth order Runge-Kutta (RK-4) and second order Runge-Kutta (RK-2) by our TEPC method with Lorenz oscillators through a RF classifier.} 
\label{results}
\end{figure}

As another proof of principle, we demonstrate quantitative ML predictions of molecular properties from chaotic systems.  
To this end, we consider protein B-factor predictions. Protein B-Factor or the Debye-Waller factor is used in protein crystallography to describe the attenuation of X-ray  caused by thermal motion. The analysis of protein B-factors sheds light on the large scale and long time functional behaviors of native state macromolecules  \cite{haliloglu1997gaussian,park2013coarse}. There are many existing methods for protein flexibility analysis. For example, Park et al. \cite{park2013coarse} compared the performance of normal mode analysis (NMA) \cite{go1983dynamics} and  Gaussian network model (GNM) \cite{haliloglu1997gaussian} for three sets of proteins. Among them, the GNM is the most popular method for protein flexibility analysis \cite{haliloglu1997gaussian}. Xia et al. developed generalized  GNM and generalized anisotropic network model (ANM) for protein B-factor predictions \cite{xia2015multiscale}. Opron et al. proposed a multiscale flexibility rigidity index (FRI) model to achieve nearly  20\% improvement to the GNM prediction\cite{opron2015communication}. 
Recently, Cang et al. introduced an evolutionary homology method  to analyze the protein flexibility and obtained the state-of-the-art performance \cite{cang2020evolutionary}.  

In the present work, we utilize a few benchmark datasets that have been intensively studied in 
the literature   \cite{park2013coarse,opron2014fast,cang2020evolutionary}, and thus, the performance of the proposed methods can be compared. The largest dataset involves 364 distinct proteins, having tens to thousands of amino acid residues \cite{opron2014fast}.

In the proposed approach, each protein residue is represented by its C$_\alpha$ atom. Each C$_\alpha$ atom is assigned with a nonlinear oscillator. We choose the Lorenz system with parameters $\alpha=10$, $\gamma=60$,  $\beta=8/3$, and $\epsilon=13.6$ with the connectivity matrix   being  $\Gamma={\bf I}$ (the identity matrix). The interaction strength between each pair of  C$_\alpha$ atoms is computed   with the generalized exponential function Eq.(\ref{eq2})  with $\kappa =1$ and $\sigma=3$. We further use persistent Laplacians to obtain a family of coupling matrices among the residues at various scales according to filtration. At each scale, we have a coupled chaotic system.  To generate the time trajectory of a specific oscillator (residue), we choose its initial value to one and those of other oscillators to zero in a perturbative time integration. Then, the statistic values of the trajectory are used as the inputs in the ML study of the residue's B-factor. Persistent Laplacians enable us to collect multiple sets of the statistic values for each residue. 
  
We carry out B-factor predictions by a multiple linear regression model that has been  used in  earlier papers \cite{haliloglu1997gaussian,park2013coarse,opron2014fast} and compute the average Pearson correlation coefficients (PCCs) between the predicted and the experimental B-factors. Figs. \ref{results}b and c show a comparison of average PCCs over different datasets with other existing methods, including EH \cite{cang2020evolutionary}, opFRI \cite{opron2014fast}, pfFRI \cite{opron2014fast}, MND \cite{xia2014molecular}, GNM \cite{xia2014molecular, park2013coarse}, NMA \cite{park2013coarse}, mFRI \cite{opron2015communication}, mGNM \cite{xia2015multiscale}, and mANM \cite{xia2015multiscale}. Here, TEPC (Lorenz) and TEPC (Rossler) are  the present methods implemented through the Lorenz oscillator and Rossler oscillator, respectively. We find that our method TEPC (Lorenz) outperforms all other methods in computational biophysics and achieves the largest average PCCs over all different protein datasets. Specifically, in Fig. \ref{results}b,  for three sets of proteins with small (33 samples), medium (36 samples), and large (35 samples) sizes, our average PCCs are 0.940, 0.826, and 0.789 by TEPC (Lorenz), which are improved up to 95.8\%, 71.4\%, and 59.7\% compared to those of 0.480, 0.482, and 0.494 in Ref. \cite{park2013coarse}, respectively. Additionally, in Fig. \ref{results}c, compared with the existing best results, the average PCCs 0.794, 0.796, and 0.817 obtained by TEPC (Lorenz) are about 13.8\%, 24.0\%, and 49.6\% higher than 0.698 \cite{cang2020evolutionary}, 0.642 \cite{xia2015multiscale}, and 0.546 \cite{xia2015multiscale} on 364, 362, and 300 datasets, respectively.  The potential reason for this excellent performance is that the proposed method provides  richer information of residue interactions through multiscale topological  Laplacians. More details about the results in Fig. \ref{results}b and c, and the prediction result of B-factor on 364 datasets can be found in Tables S12 and S17 of the Supporting Information, respectively. Three examples of B-factor predictions of proteins 1FF4, 2WUL, and 2Y7L with correlation coefficients 0.963, 0.955, and 0.891 are given in the Supporting Information, which suggests that our prediction results agree well with experiments. 

Figs. \ref{results}d and e show another B-factor predictions on protein calmodulin (PDB ID: 1CLL). In Fig. \ref{results}d, our TEPC method with the Lorenz oscillator displays excellent performance, and its results match closely with experiments for residues 60-90. In contrast, the GNM method demonstrates poor predictions. This difference is obvious when comparing the color representations based on B-factor values of the 1CLL protein structure in Fig. \ref{results}e.

These protein B-factor predictions not only demonstrate the usefulness of the present TEPC method in handling proteins with different sizes and different complex structures, but also indicate that the proposed approach has great potential for other important biophysics applications, such as the predictions of protein-ligand binding affinities and protein-protein interactions.

\subsection{Predicting single cell properties from chaotic systems} 

Having demonstrated the utility of topology-enabled molecular property predictions from chaos, we further demonstrate the topology-enabled cellular property predictions from chaos. Specifically, we consider single-cell RNA sequencing  (scRNA-seq)  data analysis, which is a computational process aimed at extracting meaningful biological insights from single-cell gene expression data. In the present work, we employ 13 scRNA-seq datasets as shown in Table S4 of the Supporting Information. These datasets involve tens of thousands of single cell samples collected from human or mouse. The single cells contain thousands or tens of thousands of genes. The problem is to classify these cells into various cell types, from 4 to 14 types, in various datasets based on their gene expressions.  
  
We first compute the Euclidean distance between cells and then set $\kappa =1$ and $\sigma=3$ for the exponential kernel in Eq. (\ref{eq2}) to evaluate pairwise correlations.  We employ persistent Laplacians to create a family of correlation matrices for various chaotic systems. The trajectories from chaotic systems are collected, and their statistical values are computed for cell classification.   We use 5-fold cross-validation with 10 random seeds and the random forest (RF) algorithm to predict cell type. The default parameters of RF from Scikit-learn's packages were used in the implementation. Here, we compare the overall classification performance of our method (TEPC with Lorenz oscillators) with widely used principal components analysis (PCA) method in Fig. \ref{results}f. 
	
	For PCA, we take the average value of balanced accuracy (BA) over different dimensions, namely 50, 100, 150, 200, 250, and 300. For each reduction, 20 random initializations were used to ensure robustness and stability. Additionally, before the dimensionality reduction, a log-transform was performed on the data sets. The 5-fold cross-validation is employed to assess the performance of various methods. Since scRNA-seq datasets usually have an unbalanced number of cells per type, we use a balanced accuracy score rather than the traditional accuracy score, whose details are given in Section S4 of the Supporting Information. From Fig. \ref{results}f, we find that our TEPC with Lorenz oscillators outperforms PCA on all 13 scRNA-seq dataset. Specifically, the highest improvement of BA value is up to 50.79\% on GSE45719 data and the average improvement is 11.29\% for each data set, suggesting  that TEPC is a useful tool in the classification of scRNA-seq data. The details of Fig. \ref{results}f can be found in Table S4 of the Supporting Information.

\subsection{Predicting image properties from chaotic learning} 

To further demonstrate the predictive power and robustness of the proposed topology-enabled predictions from chaos, we consider a non-biological problem. Image classification is a general task in data science and has wide-spread applications in a wide variety of fields and disciplines. In this work, we consider Coil-20, a benchmark dataset of image classifications with 1440 images \cite{nene1996columbia}. Each image has a size of 128*128, where 20 objects or classes are captured at 72 angles, and was treated as a vector of length 16384. The details of Coil-20 data are given in SectionS7 of the Supporting Information.  In our implementation, we first reduce the dimension from 16384 to 3 with uniform manifold approximation and projection (UMAP). However, t-distributed stochastic neighbor embedding (t-SNE) or PCA could be similarly used for dimension reduction. The pairwise correlations among images in the dataset are computed with the exponential kernel $\kappa =1$ and $\sigma=3$ from the three-dimensional (3D) UMAP results. Then, we use persistent Laplacians to generate a family of correlation matrices for 1440 images. For each correlation matrix, we install a chaotic system and compute a perturbative trajectory for each image. The statistical values of trajectories are used for downstream ML predictions.        
As a comparison, we also directly use the features from the UMAP, PCA, and t-SNE dimension reduction for downstream classifiers. We use three different classifiers, including random forest (RF), gradient boosting decision tree (GBDT), and k-nearest neighbor classifier (KNN) in the classification with all default parameters of Scikit-learn packages. We use the 5-fold cross-validation to evaluate the performance of classifications. All details of parameters used in Fig. \ref{results} can be found in Table S2 of the Supporting Information.

In Fig. \ref{results}g, we compare the balanced accuracy of classification with these four methods, i.e., TEPC,  UMAP, PCA, and t-SNE, where TEPC (Lorenz) and TEPC (Rossler) indicate the use of two different chaotic dynamics, i.e., Lorenz and Rossler oscillators. We find that TEPC (Lorenz) outperforms UMAP, PCA, and t-SNE methods on all three classifiers. Compared with UMAP, PCA, and t-SNE, TEPC (Lorenz) has improved the balanced accuracy up to 11.59\%, 13.53\%, and 11.64\% for RF, GBDT, and KNN classifiers, respectively. The details of Fig. \ref{results}g can be found in Table S7 of the Supporting Information.
It is interesting to note that although TEPC (Lorenz) started with the same UMAP information, it outperforms UMAP by 8.65\%, 8.75\%, and 11.14\% for RF, GBDT, and KNN classifiers, respectively. 
This amazing improvement is due to the fact that the proposed topology-enabled chaotic embedding captures the network interactions and multiscale correlations among oscillators (images). In contrast, the UMAP information is relatively isolated for each image and neglects crucial `spatial' correlations among images.

\section{Discussion}

\subsection{Robustness with respect to dynamical systems and ML models}

In order to demonstrate the robustness of the proposed chaotic learning with respect to the choice of chaotic dynamics and ML algorithms, we consider two oscillators, i.e., Lorenz and Rossler and many ML algorithms, including KNN, RF, GBDT, RNN, and LSTM. Specifically,   
we integrate the Rossler equation using the fourth-order Runge-Kutta scheme and set their parameters to $a=b=0.1$, $c=4$, and $a=35,b=3,c=28$, respectively. In Fig. \ref{results}, TEPC (Rossler) means the TEPC method with Rossler oscillators.

In Fig. \ref{results}a, TEPC (Rossler) performances as well as TEPC (Lorenz) on the EEG classification task obtaining all 100\% on three metrics including accuracy, sensitivity, and specificity, which are equivalent or higher than those of other existing methods. 

In Fig. \ref{results}b, EH indicates the evolutionary homology method, and NMA refers to the normal mode analysis. Regarding the protein B-factor prediction, for the set of small-size proteins with 33 samples, TEPC (Rossler) method with an average PCC 0.934  outperforms other existing methods. Specifically, the average PCC value is improved  20.81\% and 94.58\% compared with 0.773 and 0.480 obtained by EH and NMA methods in references \cite{cang2020evolutionary} and \cite{park2013coarse}, respectively. For the set of medium-size proteins with 36 samples, TEPC (Rossler) raises the average PCC value up to 6.17\% and 60.58\%  from 0.729, 0.482 to 0.774 compared to those of EH  method \cite{cang2020evolutionary} and reference \cite{park2013coarse}, respectively. Similarly, for the set of large-size  proteins with 35 samples, TEPC (Rossler) raises the average PCC value up to 5.4\%, 41.9\% compared to those of EH method \cite{cang2020evolutionary} and reference \cite{park2013coarse}, respectively. Additionally, in Fig. \ref{results}c,   TEPC (Rossler)  also achieves better prediction result than other existing methods in different sets of proteins with 300, 362, and 364 samples, respectively. That is,  TEPC (Rossler)   gets the average PCCs 0.748, 0.722, and 0.72 with the increment up to 37.0\%, 12.46\%, and 27.43\% compared to those of 0.546, 0.642, and 0.565 by mANM \cite{xia2015multiscale}, mGNM \cite{xia2015multiscale}, and GNM methods \cite{opron2014fast} with 300, 362, and 364 datasets, respectively. 

In Fig. \ref{results}g, for image classification task, TEPC (Rossler) obtains a comparable performance with TEPC (Lorenz) method, except with KNN classifier. The balanced accuracy values of prediction, 0.915 and 0.894 with RF and GBDT classifiers attained by TEPC (Rossler) are improved up to 2.8\% and 3.35\% compared to PCA method, and 2.92\% and 0.45\% with t-SNE method, respectively, which is similar with those of UMAP method. However, TEPC (Rossler) with KNN classifier performs worse than other methods. 

From Figs. \ref{results}a, b, c, g, though the TEPC method with different kinds of oscillators have various performances on brain EEG data, protein B-factor prediction, and image classification task, they outperform the existing advanced methods or the traditional baseline methods, like PCA or UMAP or t-SNE in most of cases, which verifies the generalizability of TEPC method is independent of specific chaotic systems. 

Finally, in Fig. \ref{results}a, TEPC (RNN) and TEPC (LSTM) were paired with Lorenz, and
TEPC (Lorenz) and TEPC (Rossler) were combined with KNN. All of these methods achieved 100\% in accuracy, sensitivity, and specificity for EEG classification, confirming that the proposed TEPC method is highly robust.  

\subsection{Impact of numerical integrators}

In the implementation of TEPC method, we all utilize the fourth-order Runge-Kutta (RK-4) method to solve Lorenz (Rossler) equations. Here, in order to discuss the effect caused by different numerical integrators on ML predictions, we show the comparison results between RK-4 and RK-2 methods on 13 single cell RNA sequencing data in Fig. \ref{results}h, where the balanced accuracy (BA) are compared by the TEPC method with the Lorenz oscillator integrating with three different classifiers, including RF, GBDT, and SVM. 

From Fig. \ref{results}h, the balanced accuracy (BA) values by RK-4 method are larger than those by Rk-2 with RF classifier on all 13 single cell data sets. For GBDT and SVM classifier, the comparison results between RK-4 method and RK-2 method can be found in Fig. S5 of the Supporting Information. These results suggest that the RK-4 method used in our numerical experiments is a robust approach to provide reliable and efficient prediction results. The details of Fig. \ref{results}h can be found in Table S6 of the Supporting Information.

\subsection{Impact of model parametrization }
To analyze the best parameter for exponential function of Eq. (\ref{eq2}), we test a range of parameters to predict the B-factor averaged on 6RXN protein in Fig. S1 of the Supporting Information, where the best average B-factor is got around $\kappa =1$ and $\sigma=3$. This result suggests that the TEPC model introduced in present study can be a parameter-free model by setting $\kappa =1$ and $\sigma=3$.  Our findings do not depend on chaotic models and model parametrization.

\subsection{Residue-Similarity (R-S) plots}
In this section, we implement the residue-similarity (R-S) analysis \cite{hozumi2023preprocessing} for the visualization of classification performance on single cell RNA sequencing (scRNA-seq) datasets. In classification tasks involving two classes, conventional approaches such as Receiver Operating Characteristic (ROC) curves and Area Under the ROC Curve (AUC) serve comparable purposes to R-S scores. However, unlike these traditional methods, R-S scores possess the capability to be extended to scenarios with any number of classes, which is advantageous over the traditional methods.

\begin{figure}[!tpb]
\centering
\includegraphics[width=16.cm]{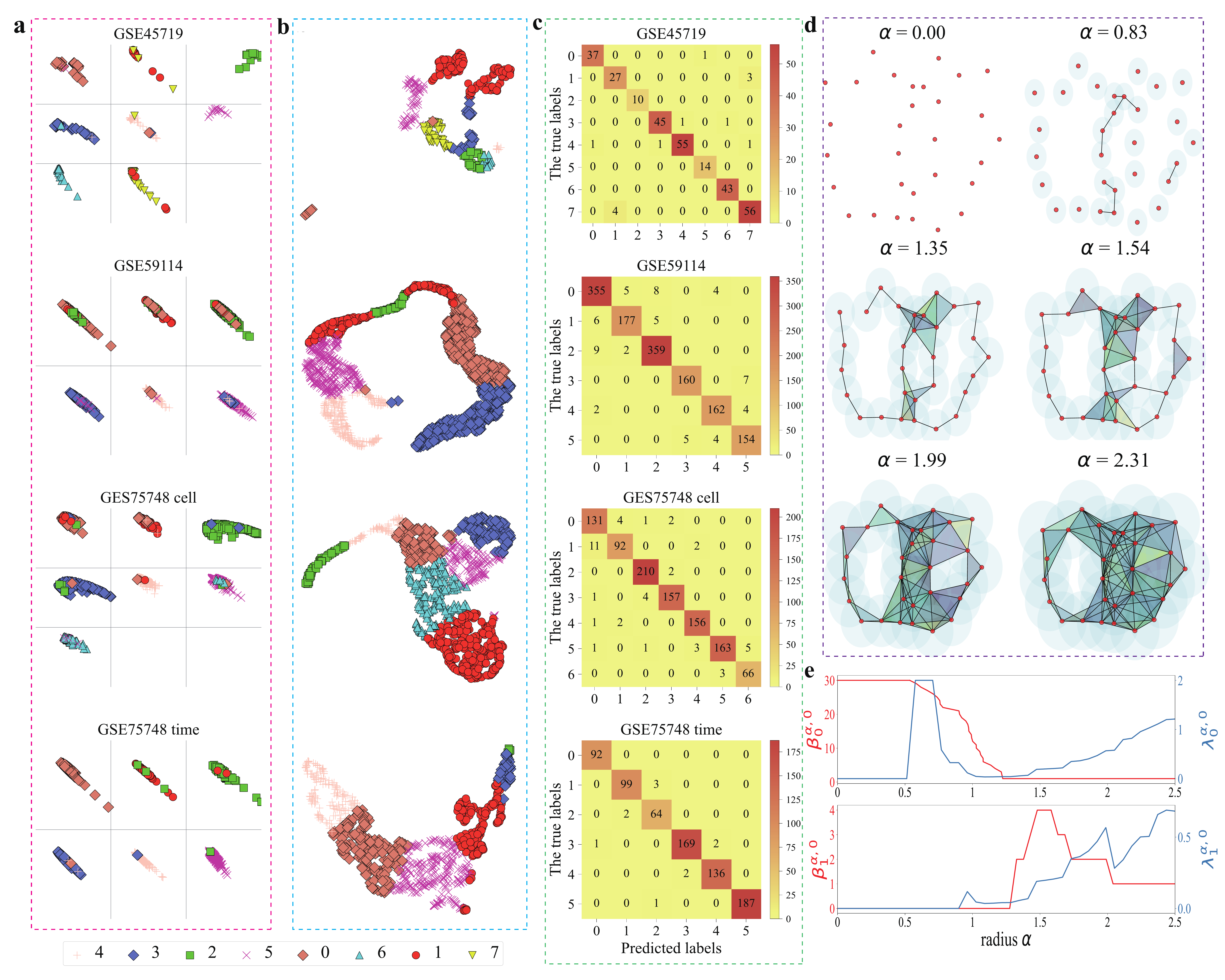}
\caption{R-S plots, UMAP plots, and the confusion matrix of RS plots of four scRNA-seq data sets, including GSE45719, GSE59114, GSE75748 cell, and GSE75748 time data sets. $\textbf{a}$: The R-S plots. The TEPC method was used to reduce the scRNA-seq data sets to 30 components with $\kappa =1$ and $\sigma=3$. 5-Fold cross-validation was used to split the data into five parts, where four parts were used for training and one part was used for testing the random forest (RF) classifier. The RS score was computed for the testing set, and all 5 folds were visualized. Each section corresponds to 1 of the 8 true cell types, and the sample’s color and marker correspond to the predicted label from the RF classifier. $\textbf{b}$: The UMAP plots. The data was log-transformed and any genes with less than $10^{-6}$ variance were removed before applying the reduction. Samples were colored according to their cell types. $\textbf{c}$: The confusion matrix based on the classification results of R-S plots.  $\textbf{d}$: The illustration of filtration process of point clouds with the generation of a series of simplicial complexes. $\textbf{e}$ gives the changes of persistent Betti numbers $\beta^{\alpha,0}_0$, $\beta^{\alpha,0}_1$, and the smallest nonzero eigenvalues of persistent Laplacian $\lambda^{\alpha,0}_0$, $\lambda^{\alpha,0}_1$ with the increasing of filtration radius $\alpha$, respectively. }
\label{singlecell_clustering}
\end{figure}

Figs. \ref{singlecell_clustering}a-b show a comparison between the R-S plot and two-dimension (2D) plots of UMAP for four scRNA-seq data sets, including GSE45719, GSE59114, GSE75748 cell, and GSE75748 time. For each data set, it was partitioned into five subsets using 5-fold cross-validation. Four subsets were utilized to train the random forest (RF) classifier, while one subset was reserved for testing. Subsequently, residue and similarity scores were calculated for each sample and visualized based on their true cell type. Samples were then color-coded according to the labels predicted by the RF classifier. The $x$-axis represents residue scores, while the $y$-axis represents similarity scores. Both scores range from 0 to 1, with 1 indicating optimal performance. A well-separated and clustered reduction is indicated by the top-right corner of the plot. The TEPC model was used to reduce the dimension to 30 supergenes with $\kappa =1$ and $\sigma=3$.  

Fig. \ref{singlecell_clustering}c represents the confusion matrix of R-S plots, where the $x$-axis means the true labels and the $y$-axis means the predicted labels. For instance, for the GSE45719 data set, in the first row of figure, we can find that classes 2, 5, and 6 are all classified correctly, and classes 0, 1, 3, 4, and 7 are classified correctly except one, three, three, four samples, respectively. However, in UMAP plots, the samples of classes 0, 1, 3, and 7 do not form compact individual cluster, and there are two obvious separated clusters for each class. These findings suggest that although UMAP has a good clustering visualization, an R-S plot has  better prediction performance than UMAP. Similar results are found in the cases of GSE59114, GSE75748 cell, and GSE75748 time data sets, shown in the last three rows of Figs. \ref{singlecell_clustering}a-c. The clustering visualization of left scRNA-seq data sets is given in Figs.S2, S3, and S4 of the Supporting Information.

\subsection{Applications of chaotic learning to other fields} 

In present work, chaotic learning has been applied to diverse complex systems, 
from single molecules (proteins), single cells, organs (brain EEG),  to non-biological data (images). 
While all examples shown in this work are about element-specific predictions, the proposed TEPC can be used for global property predictions by appropriately integrating element-specific features into global ones, such as features for a whole protein, which can be used for global predictions, such as protein solubility, toxicity, folding free energy, etc. 
 
Here, by integrating the ML, chaos theory, and persistent Laplacian, we propose a chaotic learning paradigm to predict the properties of  general  data  from unpredictable chaotic dynamics. Our paradigm can be used not only as a tool in ML predictions, but also as a new method for the analysis of times series, such as the brain EEG data. 

The proposed chaotic learning  can be improved by other topological Laplacian methods, such as  persistent hyerdigraph Laplacian and persistent sheaf Laplacian \cite{wei2025persistent},   as well as better time integrators. 
 
\section{Methods}
We consider a spatial network or high-order hypergraph where the nodes mean objects or particles, neurons and the links or edges between nodes represent the interaction between nodes. The distance between two nodes is directly associated with their interaction strength. Additionally, we assume that all functions of nodes are solely determined by the network structure. Therefore, functional predictions can be carried out with the structural information without resorting to the ultimate interactions. We build up a connectivity matrix from the interactive network, and we represent the dynamics of each node by a nonlinear oscillator, and couple nodes with the connectivity matrix extracted from the data. Finally, we review persistent Laplacian and propose a new filtration method, Laplacian filtration, instead of a conventional filtration method. In this section, we discuss the details of our main methods, mapping interaction to topology and persistent Laplacian analysis. A brief discussion of stability analysis of coupled chaotic systems and the details of residue-similarity scores are given in SectionsS1.2 and S9 of the Supporting Information, respectively. 

\subsection{Mapping interaction to topology}
Topological relations or connectivities among network nodes are basic ingredients in network nonlinear dynamics models. As discussed above, this topological information can be extracted from the interaction of a given node of interest in an $n$-dimensional space. Let us consider a network of $N$ nodes located on different regions 
${\bold{r}_1},{\bold{r}_2}, \cdots ,{\bold{r}_N}$, ($\bold{r}_j \in {\mathbb{R}^n}$), where nodes can be neurons, atoms, amino acids residues in a molecule, or people in society if applied in other scenarios. The distance between the $i$th and $j$th nodes is given by $d_{ij}(r_i,r_j)=
\left\| {{\bold{r}_i} - {\bold{r}_j}} \right\|_2$. The connectivity matrix must agree with the driven and response relation between two dynamics systems. Moreover, we suppose that all nodes in the network are mutually linked and their interactions decay as a function of their distance $\bold{A}_{ij}(d_{ij})$. The simplest form of the connectivity matrix is the Kirchoff (or connectivity) matrix generated by cutoff distance $\sigma_{ij}$:
\begin{equation}\label{eq1}
{A_{ij}} = \left\{ {\begin{array}{*{20}{c}}
{1, \quad {\rm{ }}\forall {d_{ij}} \le {\sigma _{ij}},i \ne j}\\
{0,\quad {\rm{    }}\forall {d_{ij}} > {\sigma _{ij}},i \ne j}\\
{ - \sum\nolimits_{j \ne i} {{A_{ij}},} \quad {\rm{  }}\forall i = j.}
\end{array}} \right.
\end{equation}

In order to take the distance effect in a more practical manner into the consideration, smooth and monotonically decreasing radial basis functions or the delta sequence kernels of positive type are often employed for convenience \cite{wei2000wavelets}. Here, generalized exponential functions 
\begin{equation}\label{eq2}
{A_{ij}} = \left\{ {\begin{array}{*{20}{c}}
{{e^{{{ - d_{ij}^k} \mathord{\left/
 {\vphantom {{ - d_{ij}^k} {k\sigma _{ij}^k}}} \right.
 \kern-\nulldelimiterspace} {k\sigma _{ij}^k}}}}, \quad {\rm{  }}\forall i \ne j,{\rm{  }}k = 1,2, \cdots }\\
{ - \sum\nolimits_{j \ne i} {{A_{ij}},} \quad {\rm{    }}\forall i = j}
\end{array}} \right.   
\end{equation}
and power law functions are considered,
\begin{equation}\label{eq3}
    {A_{ij}} = \left\{ {\begin{array}{*{20}{c}}
{{{\left( {{{{d_{ij}}} \mathord{\left/
 {\vphantom {{{d_{ij}}} {{\sigma _{ij}}}}} \right. 
 \kern-\nulldelimiterspace} {{\sigma _{ij}}}}} \right)}^{ - \upsilon }},\quad {\rm{   }}\forall i \ne j,{\rm{ }}\upsilon  > 1{\rm{ }}}\\
{ - \sum\nolimits_{j \ne i} {{A_{ij}},} \quad {\rm{            }}\forall i = j,}
\end{array}} \right.
\end{equation}
where $\sigma_{ij}$ are characteristic distance between nodes. In the present work, $\sigma_{ij}$ is a tuneable parameter. Note that the above matrix functions have already been applied to other methods, like flexibility analysis methods in biology system \cite{bahar1998vibrational,atilgan2001anisotropy}. But, the present construction of these functional forms is based on the driven and response relationship of coupled dynamical systems \cite{pecora1997fundamentals}.

Formulas (\ref{eq1})-(\ref{eq3}) can be used to map the geometry of a network into topological relations or connectivities. The connectivity matrix $\bf A$ is an $N \times N$ symmetric, diagonally dominant matrix, and the elements in the matrix are not interaction potentials among nodes. For simplify, we set that the characteristic distances of all nodes are the same $\sigma_{ij}= \sigma$.

\subsection{Persistent Laplacian analysis}

Persistent homology (PH) is a new branch of algebraic topology and has been developed as a new multiscale representation of topological features applied successfully in various fields, including mathematics \cite{edelsbrunner2008persistent}, chemistry \cite{townsend2020representation}, 
etc. It contains a filtration process to generate a family of topological spaces so that the system can be analyzed at multiscales.  More discussion of PH theory is given in Section S8 of the Supporting Information. However, PH cannot detect the homotopic shape evolution of the system during filtration. Recently, persistent Laplacian (PL) or persistent spectral graph (PSG) is developed to fill the gap. PL not only retains the full set of topological invariants in their harmonic spectra as PH does but also captures the homotopic shape evolution of data during the filtration in their non-harmonic spectra that PH does not describe \cite{wang2020persistent}.
PL is a special case of persistent topological Lpalacians, including the persistent Hodge Laplacian defined on a Riemannian manifold \cite{chen2021evolutionary}, and 
 a variety of other formulations defined on many topological spaces, such as cellular sheaf, hyperdigraphs, path complexes, etc. 
 Like PH, PLs utilize a filtration process  to generate a series of geometric shapes and associated topological spaces, on which persistent spectral graphs are defined.   The change in the null space dimensions of the PLs during the filtration indicates the persistence of topological invariants, while the nonzero eigenvalues and related eigenfunctions of the PLs reveal the geometric shape evolution of the data during the filtration.

An oriented simplicial complex $K$ is a sequence of subcomplexes $(K_t)_{t=0}^m$ of $K$:
\begin{equation}
    \emptyset = K_0 \subseteq K_1 \subseteq K_2 \subseteq \cdots \subseteq K_m = K. 
\end{equation}
Considering the corresponding chain group of each subcomplex $K_t$ and $q$-boundary operator: $C_q(K_t)$ and $\partial_q^t: C_q(K_t) \to C_{q-1}(K_t)$, if $0<q \le \dim K_t$, we have 
\begin{equation}
    \partial_q^t(\sigma_q) = \sum_{i}^q(-1)^i \sigma^i_{q-1}, \quad \forall \sigma_q \in K_t,
\end{equation}
where $\sigma_q = [v_0, \cdots, v_q]$ is any $q$-simplex and $\sigma^{i}_{q-1} = [v_0, \cdots, \hat{v_i} ,\cdots,v_q]$ the oriented $(q\!-\!1)$-simplex constructed by removing $v_i$. It is important to define an adjoint operator of $\partial_q^t$ as the coboundary operator $\partial_q^{t^{\ast}}: C^{q-1}(K_t) \to C^q(K_t)$,   mapping from $C_{q-1}(K_t)$ to $C_q(K_t)$ by  the isomorphism   between cochain groups and chain groups $C^q(K_t)\cong C_q(K_t)$.

For simplicity, denoting $C_q^t$ the chain group $C_q(K_t)$ and using the natural inclusion map from $C_{q-1}^t$ to $C_{q-1}^{t+p}$, one defines the subset of $C_q^{t+p}$ as $\mathbb{C}_q^{t,p}$ with the boundary being in $C_{q-1}^t$ as
\begin{equation}
    \mathbb{C}_q^{t,p} \coloneqq \{ \beta \in C_q^{t+p} \mid \ \partial_q^{t+p}(\beta) \in C_{q-1}^{t}\}.
\end{equation}
On this subset, we denote the $p$-persistent $q$-boundary operator by $\eth_q^{t,p}: \mathbb{C}_q^{t,p} \to  C_{q-1}^{t}$and the corresponding adjoint operator by $(\eth_q^{t,p})^{\ast}: C_{q-1}^{t}  \to  \mathbb{C}_q^{t,p}$ via the determination of cochains with chains.  We can define the $q$-order $p$-persistent Laplacian operator $\Delta_q^{t,p}: C_q^t \to C_q^t$ in the filtration \cite{wang2020persistent}
\begin{equation}\label{eq14}
    \Delta_q^{t,p} = \eth_{q+1}^{t,p} \left(\eth_{q+1}^{t,p}\right)^\ast + \partial_q^{t^\ast} \partial_q^t.
\end{equation}
The matrix representation of $\Delta_q^{t,p}$ in the simplicial basis is 
\begin{equation}
    \mathcal{L}_q^{t,p} = \mathcal{B}_{q+1}^{t,p} (\mathcal{B}_{q+1}^{t,p})^T + (\mathcal{B}_{q}^t)^T \mathcal{B}_{q}^t.
\end{equation}
The topological invariants of the corresponding persistent homology defined by the same filtration can be recovered from the kernel of PL Eq. (\ref{eq14}):
\begin{equation}
    \begin{aligned}
        \beta_q^{t,p} = \dim \ker \partial_q^t - \dim \im \eth_{q+1}^{t,p} = \dim \ker \mathcal{L}_q^{t,p} = \# \text{0 eigenvalues of } \mathcal{L}_q^{t,p}.
    \end{aligned}
\end{equation}
In this study, we only focus on the 0, 1, and 2nd-order PLs for brain neural network data. Mathematically,  $\beta_q^{t,p}$ from the null space of LPs tracks the number of independent $q$-dimensional holes in $K_t$ that are still alive in $K_{t+p}$. Therefore, it gives the same topological information as PH does. However, the non-zero eigenvalues of the PLs reveal the homotopic shape evolution of data during filtration.  

It is worth noting that rather than conventional filtration method directly by studying the simplicial complexes themselves in PH or PL, in the present work, we implement a Laplacian filtration procedure on our weighted Laplacian matrix to construct a family a subgraphs based on an accumulation threshold. The function of the weighted Laplacian matrix is as following:

 \begin{equation}
    \begin{split}
        L=(l_{ij}), l_{ij}= \left \{
        \begin{array}{cc}
           l_{ij},  &  i \neq j, i,j=1,\dots ,n \\
          l_{ii} = -\sum_{j=1}^n l_{ij}.  & 
        \end{array}
        \right.
    \end{split}
 \end{equation}

For $i \neq j$, let $l_{\text{max}}=\text{max}(l_{ij})$, $l_{\text{min}}=\text{min}(l_{ij})$, and $d=l_{\text{max}}-l_{\text{min}}$. We set the $k$th Persistent Laplacian $L^k,k=1,\dots,p$:
 \begin{equation}
    \begin{split}
        L^k=(l^k_{ij}), l^k_{ij}= \left \{
        \begin{array}{cc}
           0,  &  \text{if} \  l_{ij} \leq (k/p)d+l_{\text{min}} \\
          -1 & \text{otherwise} \\
          l_{ii}^k=-\sum_{j=1}^n l_{ij}^k. &\\
        \end{array}
        \right.
    \end{split}
 \end{equation}

Additionally, Fig. \ref{singlecell_clustering}d describes the topological representation of point clouds through Vietoris-Rips complexes with different filtration radius $\alpha$. The behaviors of Betti-0 and Betti-1 of persistent homology over filtration are illustrated as the red lines in Fig. \ref{singlecell_clustering}e. In contrast, the lowest nonzero eigenvalues ($\lambda^{\alpha,0}_0$ and $\lambda^{\alpha,0}_1$) of persistent Laplacian over the same filtration are illustrated by blue lines in Fig. \ref{singlecell_clustering}e. Persistent Laplacian captures not only all topological changes but also homotopic geometric evolution during the filtration. The topological representation of point clouds in regular octagon can be found in  the Supporting Information.

\section*{Data and code availability}
The related codes studied in the present work are available at: \url{https://github.com/kelu0124/TEPC/tree/main}.

\section*{Acknowledgment}
This work was supported in part by NIH grants R01GM126189 and R01AI164266, NSF grants DMS-2052983, DMS-1761320, and IIS-1900473, NASA grant 80NSSC21M0023, MSU Foundation, Bristol-Myers Squibb 65109, and Pfizer. The work of Huahai Qiu and Bengong Zhang was supported by the National Natural Science Foundation of China under Grant No.12271416 and No.12371500, respectively.

\section*{Competing interests}
The authors declare no competing interests.

\end{document}